\documentstyle[12pt]{article}
\def\eq{\begin{equation}}
\def\en{\end{equation}}
\newcommand \be  {\begin{equation}}
\newcommand \bea {\begin{eqnarray} \nonumber }
\newcommand \ee  {\end{equation}}
\newcommand \eea {\end{eqnarray}}

\def\B{\beta}

\def \bi{\bibitem}

\def \bs{ {\bf s}}

\def\d{{\rm d}}
 \def\(({\left(}
 \def\)){\right)}
\def\bi{\bibitem}
\def\jpa{J.Phys.A}

\def\hs{\hat{s}}

\def \tp{\tilde{p}}
\def \ov{\over}
\def\bm{{\bf m}}
\def \a{\alpha}
\def \b{\beta}

\def\bQ{{\bf Q}}
\def \d{{\rm d}}
\def \e{{\rm e}}
\def \del{\delta}
\def \nn{\nonumber}
\def \beqna{\begin{eqnarray}}
\def \eeqna{\end{eqnarray}}
\def \beq{\begin{equation}}
\def \eeq{\end{equation}}
\def \be{\begin{equation}}
\def \ee{\end{equation}}

\def \ov{\over}

\def \ol{\overline}
\def \a{\alpha}

\def \b{\beta}
\def \r{\right}

\def \l{\left}

\def \la{\langle}
\def \ra{\rangle}
\def \Tr{{\rm Tr}}
\def \eps{\epsilon}
\def \ab2{\alpha\beta^2}
\def \s{\sigma}
\def \la{\langle}
\def \ra{\rangle}

\def \cD{{\cal D}}

\def \pt{\tilde p}
\newcommand{\pub}[4]{{\em #1 }{\bf #2}, #3 (#4)}

\begin{document}

\baselineskip25pt

\begin{center}

{\LARGE\bf Recipes for metastable states in Spin Glasses}
\vskip1cm

\end{center}

\baselineskip20pt

\begin{center}

{\Large Silvio Franz (1), Giorgio Parisi (2)}
\vskip1cm

{\large
(1)  NORDITA and CONNECT\\
     Blegdamsvej 17,\\
     DK-2100 Copenhagen \O\\
     Denmark\\
e-mail: {\it franz@nordita.dk }\\ }

\vskip0.5cm
{\large
(2)  Dipartimento  di Fisica \\
Universit\`a di Roma I, ``La Sapienza''\\
P.le A.Moro 5\\
00185 Rome\\
Italy }
\vskip0.5cm

{\large March 1995}
\vskip1cm

\end{center}

\baselineskip16pt

\begin{flushleft}
cond-mat/9503167
\end{flushleft}


{\bf Abstract:} In this paper we develop a method introduced by one
of us to study metastable states in spin glasses.
We consider a `potential function' defined as the free energy of a system
at a given temperature $T$ constrained to have a fixed overlap with a
reference configuration of equilibrium at temperature $T'$. We apply the method
to the spherical p-spin glass and to some generalization of this model
in the range of temperatures between the dynamic and the static transition.
The analysis  suggests a correspondence
among local minima of the potential and metastable states. This correspondence
is confirmed studying
the relaxation dynamics at temperature $T$
of a system starting from an initial configuration equilibrated at
a different temperature $T'$.


\vfill
\eject
\section{Introduction}
The off-equilibrium dynamics of glassy system is a fascinating subject.
Experimental \cite{agerev} and numerical
\cite{numrev}
 evidence show that out of equilibrium phenomena
persist in glasses for the largest reachable observation times.

Recently a partial comprehension
of these phenomena in mean field
has been achieved \cite{cuku,frame,cukusk}.
The phenomenon of aging is described in the mean field theory
as an asymptotic
stationary state, where time translation invariance and the fluctuation
dissipation relation do not hold. This regime turns out to be closely
related to the nature of the static glassy transition. The models studied up
to now can be divided in two classes according to their pattern of replica
symmetry breaking (RSB) \cite{mpv}. If the RSB is ``continuous'', i.e. if the
Parisi
order parameter function $q(x)$ is continuous, then the static and the dynamic
transitions occur at the same temperature.
The asymptotic state
is such that  expected values of  quantities which depend only on the
configuration of the system at a single time
 (e.g. the energy or the distribution of the magnetizations) tend to their
Boltzmann-Gibbs values. If instead
the Parisi function is discontinuous, e.g. a single step function,
the dynamic transition occurs at a temperature higer then the static one, and
the quantities mentioned above tend to limits different to their canonical
averages. However, a careful analysis in the {\it spherical
p-spin model} \cite{cuku} have shown how this limiting values can be related to
the values of the same observables in a particular class
 of metastable solutions
of the TAP equations, with free energy higher than the ones dominating
the partition function.
It then raises spontaneously the question whether  it is in general
 possible to relate the properties of the asymptotic dynamic state
to some static properties of the systems.

This paper constitute a step forward in this direction. Our basic tool
of investigation will be a ``potential function'' introduced in \cite{parpot},
 defined as the
minimal work needed to keep a system at temperature $T$ at a fixed
overlap  from a typical equilibrium configuration of the same system at
a different temperature
$T'$. The models we choose to study are a family of spherical
spin glass models which present a one
 step RSB.
The most studied representant of this class is the
spherical $p$-spin model.
This model has the remarkable (and atypical) property that the order in free
energy of the solutions of the TAP equations do not vary with the
temperature.
It is interesting to consider more general models
where the order of the solution depends of $T$, and
in the spin glass phase the low-lying states at different temperature are not
correlated.

The basic theses of this paper is that the minima of the potential are related
to (meta)stable states. To confirm this point we study the dynamics
at temperature $T$ starting from a  configuration which is perfectly
thermalized at a different temperature $T'$ at the initial time.

Work on
subjects related to the ones treated in this paper
has been very recently achieved by R.Monasson \cite{remi}. We thank him
to make us aware of some of his results prior to publication.
Applications of 'potential' model to Ising spin has also been done by
\cite{MPR5}.

The organization of the paper is the following:
In section 2 we present a short summary of the static and dynamic
properties of the model. This chapter is part review and part new elaboration.
Then in section 3 we introduce and discuss some basic properties of the
potential function. Section 4 is devoted to the study of the potential
for the spherical model. In section 5 we present the dynamic theory
of the evolution of a system at temperature $T$ starting from an equilibrium
configuration  at temperature $T'$. We finally sketch our conclusions.

\section{The Model}

 The spherical $p$-spin model \cite{pspin}
is defined by the Hamiltonian
\be
H_p[{\bf s}]=\sum_{i_1<i_2<...<i_p}J_{i_1,...,i_p}s_{i_1}...s_{i_p}
\label{mon}
\ee
where the ``spins'' $s_i$, $i=1,...,N$  are real variables  subjected
to the constraint $(1/N)\sum_i s_i^2=1$,
and the couplings $J_{i_1,...,i_p}$
are independent Gaussian variables with variance:
$\ol{J_{i_1,...,i_p}^2}={p!\ov 2 N^{p-1}}$.

It can be observed
 that such model is (dynamically and statically) equivalent to
a model in which the functional form of the  Hamiltonian is
not specified explicitly and one just take it as
a random Gaussian function of the
spin configuration with correlation function
given by
\be
\ol{ H[{\bf s}]H[{\bf s'}] }=N{1\ov 2}q_{\bf s,s'}^p
\ee
$q_{\bf s,s'}={\bf s\cdot s'}/N=(1/N)\sum_i s_i s'_i$
is the overlap among the two configurations $\bs$ and $\bs'$.
One can  then easily
generalize the spherical model
to random Hamiltonian \cite{MP,theo} with arbitrary correlation
functions:
\be
\ol{ H[{\bf s}]H[{\bf s'}] }=N f(q_{\bf s,s'}).
\ee
Polynomial functions  $f(q)$ with positive coefficients correspond to
Hamiltonians representable as
a sums of independent monomials of the kind (\ref{mon}).
Depending on the function $f$ the model can undergo
either  `continuous replica symmetry breaking' or `discontinuous'
 one.
If the function $g(q)=(f''(q))^{-3/2} f'''(q)$ is {\it monotonically
increasing }
with $q$ in the interval [0,1],
we are in the first case and we find a low temperature phase characterized
by a continuous Parisi function $q(x)$\footnote{It has been observed in
\cite{theo} that the model with $f(q)=(1/2)
(q^2+q^4)$
 has the same critical behaviour the SK spin-glass.}.
If  otherwise the function $g$ {\it monotonically decreasing}
one finds  $q(x)$ to be a single
step function in the spin glass phase.
Throughout this paper, if  not stated otherwise, we will consider this second
case; we
will also always consider functions $
f$ such that $f'(0)=0$. Here the asymptotic off-equilibrium dynamics and the
statics lead to
different results as far as the expect
ed values mentioned in the introduction
are concerned. Notably it is found for the energyat temperatures less than a
dinamical critical temperature
$T_D$:
$\lim_{t\to\infty}E_{Dyn}(t)>E_{Gibbs}$,
where $ E_{Dyn}(t)$ is the energy computed in the infinite volume limit
starting from a random initial
configuration .

For further reference we give here the result of standard static and
dynamic analysis for the generalized spherical model.

\subsection{ Statics}

{}From the standard study of the equilibrium measure of the model by means of
the replica method it is find a
 free energy functional
 given (in standard notations)
 by the one step RSB form
\beqna
F & &=-{T\ov N}\log Z
\nn\\
& & =-{1\ov 2}\l\{ \b [f(1)-(1-x)f(q)] +T\log(1-q)
-T{1\ov x}\log\l( {1-q\ov 1-(1-x)q} \r) \r\}
\eeqna
$q$ and $x$ are variational parameters with respect to
which $F$ has to be maximized.
In the high temperature phase $q$ is equal to zero, and it is
discontinuous at the transition point $T=T_S$, where $x=1$ and $q$ verifies:
\be
\b_S^2 f(q)+\log (1-q)+q=0\;\;\;\;\;\;
\b_S^2 f'(q)-{q\ov 1-q}=0.
\ee

\subsection{Off-equilibrium dynamics}

 One studies  here
the Langevin dynamics associated to the model with random initial condition.
The system is analyzed supposing that the thermodynamic limit is taken
{\it before} the infinite time limit, so as to prevent full equilibration in
presence of a phase transition.

The
relevant objects of investigation (order parameters) turn out to
be the correlation function $C(t,t')=1/N \sum_i s_i(t)s_i(t') $, and its
associated response function
$G(t,t')=1/N \sum_i {\del s_i(t) \ov \del h_i(t')}$
(here and in what follows $t\geq t'$).
One finds then the set of coupled equations:
\beqna
{\partial G(t,t') \over \partial t}&=& -\mu(t)G(t,t')
+\int_0^t ds  \ f''(C(t,s))G(t,s)(G(t,t')-G(s,t'))\ ,
\\
{\partial C(t,t') \over \partial t} &=&-\mu C(t,t')
+  \int_0^{t'} ds \  f'(C(t,s)) \ G(t',s)\nonumber\\
&+& \int_0^t ds \  f''(C(t,s))G(t,s) \ (C(t,t')-C(s,t'))\ ,
\nonumber\\
\mu(t)&=&
 \int_0^t ds  \ f'(C(t,s)) \ G(t,s) \nonumber\\
&+& \int_0^t ds \  f''(C(t,s))G(t,s) \ (C(t,t)-C(s,t))\ +T. \
\eeqna

Following the analysis of \cite{cuku} for the $p$-spin model,
one finds that for large times,
two regimes are important: a first one in which one takes the
limit $t,t'\to\infty$ fixing to a finite value the difference
$\tau=t-t'$ and a second one where the limit is taken fixing
the ratio $\lambda=h(t')/h(t)$.  $h$ is a function that at present the theory
is not able to specify.
\begin{itemize}
\item
In the first regime the correlation  function
decays with $\tau$ from the value  one to a finite value $q$ (dynamic
Edward-Anderson parameter)  and
 the fluctuation dissipation relation $TG_{as}(\tau)=-\partial C_{as}/\partial
\tau$ is satisfied.
\item
In the second regime one finds a non time homogeneous form
$C(t,t')={\cal C} (h(t')/h(t))$, with ${\cal C}(\lambda) $ monotonically
increasing with $\lambda$ and
${\cal C}(1)=q$ and ${\cal C}(0)=0$.
The response function
is  equal in this regime  to  $TG(t,t')=x {\partial C(t,t')\ov \partial t'}$.
The constant $x$ is found to be a number between zero and one in the low
temperature phase.
The parameters $q$ and $x$ are solution of the equations:
\be
x={1-q\ov q}\l[{qf''(q)\ov f'(q)}-1\r];\;\;\;\;\;\;\;
\b^2 f''(q) (1-q)^2=1.
\label{camm}
\ee
\end{itemize}

The dynamical transition is marked by the condition $x=1$ at $T=T_D$, and one
can check that $T_D>T_S$.
The first of
eq. (\ref{camm}) is equivalent to the variational equation $\partial F/\partial
q=0$
in statics, while the second one coincides with the condition that
`replicon eigenvalue' of the fluctuation matrix
in replica space is equal to zero. We will refer to that equation
as marginality condition.

\subsection{ TAP approach}

For completeness we mention a third approach that is useful to investigate the
system.
This is the one of the  TAP equations, in which
one writes mean  field equations for the magnetizations for fixed disorder.
Using the diagrammatic approach   of \cite{CS}, based on the skeleton
expansion introduced in \cite{cirano}, one easily finds the TAP free energy
function:
\be
F_{TAP}[\bm,q]={1\ov N} H[\bm]-{T\ov 2} \log (1-q) -{\b \ov
2}[f(1)-f(q)-(1-q)f'(q)]
\ee
where the variables $m_i$ represent the average magnetizations, and  the
self-overlap $q$ is given by
$1/N \sum_i m_i^2$.
The physical states are solutions of variational equations
both with respect to the $m$'s and $q$.

In the $p$-spin model it was found useful \cite{kpv}
 to rescale the variables to
\be
m_i\to \sqrt{q} s_i
\;\;\;
{1\ov N} \sum_i s_i^2 =1.
\ee
It then follows from (\ref{mon}) that
\be
H[\bm]=q^{p/2}H[\bs]
\ee
The points of extreme  of $F_{TAP}$ with respect to the $s_i$, as
well as their order in free energy do not depend on temperature \cite{kpv}.
Let us stress here that this is a highly
 non generic property which depends critically
on the
homogenity
 of the Hamiltonian (\ref{mon}). In general we can expect
the order of the solutions to depend on temperature. The situation
in which this happens for the lowest states is usually called chaotic
in the literature \cite{kondor}, and has been recently fully
demonstrated in the SK model \cite{neyfra}. A complete analysis of the TAP
equations equations is at present missing.
In section 4 we will find evidence for chaos with respect to temperature
in spherical models with the method we propose in the next section.

\section{ The Method}

We now introduce the ``potential'', our
basic tool of analysis in this paper.

We consider  an arbitrary configuration of the spins $\bs$, drawn as
a typical realization of the canonical probability distribution
at temperature $T'$, $P_{Can}[\bs]=\exp(-\b' H[\bs])/Z(T').$
WE can then compute the cost in free energy at a temperature
$T$ (in general different from $T'$)
 to keep the system at a fixed overlap
$\pt=q_{s,\s}$ with $s$; namely
\beqna
V=-{T \ov N}\log Z[\bs;\tp]-F[T];
\label{12} \\
Z[\s;\tp] =\int \d {\bf \s}\exp\l( -\b H[{\bf \s}]\r) \del\l(\pt-q_{\bf
s,\s}\r)
\label{13}
\eeqna
where $F[T]$ is the free energy without constraint. As $Z$ in (\ref{13}) is a
sum
of positive terms, $V$ in (\ref{12}) is a positive quantity.
It is reasonable to suppose   that $V$ in addition of being
self-averaging with respect to the quenched disorder in the Hamltonian,
it is also self-averaging
 with respect to the probability distribution of
the reference configuration $\bs$.
In this problem the spins $s_i$ are quenched variables on the same foot as the
random Hamiltonian itself. In this respect the present method
allows to extend and improve the analysis of ref. \cite{FPV}, where there
were considered different `real replicas' coupled in a symmetric way.

We need then, in order to compute $V$, to perform the
average over the Hamiltonian and the reference configuration $\bs$.
\be
NV=\ol{{1\ov Z[\b']}\int \d {\bf s }\exp\l( -\b' H[\bs]\r)\l[-T \log
Z[\bs;\tp]-F[T]\r]}.
\ee

This average can be done with the aid
of replica method.  We found two strategies which consistently
lead to the same results.
The first one is based on the formula
\be
NV=-T\lim_{n\to 0}\lim_{R\to 1}\ol{ {\partial\ov \partial R }
\l\{ \sum_s \exp\l( -\b' H[\bs]\r) Z[\bs;\tp]^{R-1} \r\}^n }.
\ee
One then evaluates the average for integer $R$ and $n$ and obtains
$V$ by analytic continuation.

The starting point of the second strategy is the formula
\be
NV=-T\lim_{n\to 0}\lim_{m\to 0}\ol{
 \sum_s \exp\l( -\b' H[\bs]\r)Z[\b']^{n-1} \l( {Z[\bs;\tp]^{m}-1\ov m}\r)  }
\ee
and again one performs an analytic continuation from integer $n$ and $m$.
The
use  of one procedure instead of the other is mainly a matter of taste.
We will sketch the first stages of the formal manipulations for both
the procedures, and we will treat our model with the second one.

Let us start with the first procedure.
The replicated partition function is:
\be
Z_1^{(n,R)}=\ol{\int \d {\bf s}^{a,r} \exp
\l[
   \b' \sum_{a=1}^n H[{\bf s^{1,a}}]+\b \sum_{a=1}^n\sum_{r=2}^R
 H[{\bf s^{r,a}}]
\r]
   \prod_{a=1}^n\prod_{r=2}^R \del
   \l(
      \sum_i s_i^{1,a}s_i^{r,a}-N\pt
   \r)
               }.
\ee
One can then perform the average over the distribution of the Hamiltonian,
and introduce the order parameters:
\be
\bQ_{(a,r),(b,s)}={1\ov N}\sum_i s_i^{r,a}s_i^{s,b}.
\ee
The logarithm of $Z_1^{n,R}$ divided by $N$ is then found to be
equal to the saddle point over the $Q$'s of
\be
{1\ov 2}\sum_{r,s}^{1,R}\sum_{a,b}^{1,n} \b^r\b^s f(\bQ_{(a,r),(b,s)})
+{1\ov 2}\Tr \log {\bf Q}
\ee
where $\b^r=\b'$ for $r=1$, $\b^r=\b$ for $r=2,...,n$.
As usual one needs a scheme to perform the analytic continuation
in the number of replicas.

With the second procedure, one has
\be
Z_2^{(n,m)}=\ol{\int \d {\bf s}^{a}\int \d {\bf \s}^{\a} \exp
\l[
   \b' \sum_{a=1}^n H[{\bf s^{a}}]+\b \sum_{\a=1}^m
 H[{\bf \s^{\a}}]
\r]
   \prod_{\a=1}^m \del
   \l(
      \sum_i s_i^{1}\s_i^{\a}-N\pt
   \r)
               }.
\ee

After the average over the distribution of the Hamiltonian is performed, one
introduce
 the order parameter matrices:
\beqna
Q_{ab}={1\ov N}\sum_i s_i^{a}s_i^{b}\\
R_{\a\b}={1\ov N}\sum_i \s_i^{\a}\s_i^{\b}\\
P_{a\a}={1\ov N}\sum_i s_i^{a}\s_i^{\a}
\eeqna
with $a,b=1,...,n$ and $\a,\b=1,...,m$.
Combining  the order parameters
 in the single $(n+m)\times(n+m)$ matrix
\be
\bQ=
\l(
\begin{array}{ll}
Q & P\\
P^T & R
\end{array}
\r)
\ee
one finds
\be
{1\ov N}\log Z_2^{(n,m)}=
{1\ov 2}\l[\sum_{a,b}^{1,n} \b'^2 f(Q_{a,b})+\sum_{\a,\b}^{1,n} \b^2
f(R_{\a,\b}
)+2\sum_{a,\a}^{1,n} \b\b' f(P_{a,\a})\r]
+{1\ov 2}\Tr \log {\bf Q}.
\label{25}
\ee

Observing that the constraint implies $P_{1,\a}=\pt$ for any $\a=1,...,m$,
a sensible ansatz is to assume that the matrix $P$ has elements $P_{a,\a}\equiv
P_a$ independent of $\a$ for any $a$. A simple computation based on the
formula $\Tr \log (\bQ )\sim \log\l( \int \d {\bf x} \exp(- {\bf x}\bQ{\bf x})
\r)$
reveals then the identity:
 \be
\Tr \log[\bQ]=\Tr \log [Q] +\Tr \log [R-A]
\ee
where $A$ is a $m\times m$ matrix with all the elements equal to
$A_{\a,\b}=\sum_{a,b} P_a (Q^{-1})_{ab}P_b$.

Note that with this  ansatz for $P$, $\Tr \log [Q]$ is of order $n$, while
$\Tr \log [R-A]$ is of order $m$. So, neglecting terms of order
$m$ one finds that the equation specifying the matrix $Q$ is just
\be
\b'^2f'(Q_{ab})+(Q^{-1})_{ab}=0
\ee
which is independent of $P$ and $R$ and is just the saddle point equation
for a system at equilibrium at temperature $T'$.
 This is a good consistency check
for the ansatz: the ${\bf s}$ system, which is at equilibrium, should not be
affected by the ${\bf \s}$ system.
The variational equations for $P$ and $R$ are respectively written as
\beqna
-2\b {\partial V\ov \partial P_a}&=&\b\b' f'(P_a)+
\sum_b (Q^{-1})_{ab}P_b\sum_{\a,\b}\big( (R-A)^{-1}\big)_{\a\b}=0
\label{V1}\\
-2\b {\partial V\ov \partial R_{\a\b}}&=&\b\b' f'(R_{\a\b})+
\big( (R-A)^{-1}\big)_{\a\b}=0
\label{V2}
\eeqna
For $a=1$ the l.h.s. of equation (\ref{V1}) represents
the derivative of $V$ with respect to $\tp$ and has not to be equated
 to zero.  We note that for $\tp=0$ the solution to (\ref{V1}) is simply
$P_a=0$ for any $a$. One has then ${\partial V\ov \partial \tp}=0$ and
the potential takes its minimal value $V=0$.

To solve the equations (\ref{V1},\ref{V2}) for $n,m\to 0$, it is of course
needed a continuation scheme for the various order parameter matrices.
For $Q$ the usual hierarchical ansatz \cite{mpv}
has to be employed. A sensible choice is to take also the matrix $R$ of the
hierarchical form (with
associated Parisi function $r(x)$),
and the vector with components $P_a$, as
the first line of a hierarchical matrix
$P^*_{ab}$ associated to the diagonal element $\tp$ and to a function $p(x)$.
Although this general form can be of interest for many probems, notably
whenever RSB is necessary to find the free energy,
 we will see later that a replica symmetric ansatz will be enough for the
problem addressed in this paper,
at least for the class of models we consider.

In the next section we will devote a lot of attention to the points of
minimum of the potential $V(\tp)$. The basic theses of this paper is that
the minima of the potential correspond to metastable states.

Before leaving this section we just
comment on the fact that in
  the case of
continuous RSB transition where
the probability distribution $P(q)$ is different
from zero in a whole interval $q_{min}\leq q\leq q_{max}$, we checked
that as it has to be expected,
the potential is flat and equal to zero in the whole interval
$[q_{min},q_{max}]$.

\section{ The potential in the intermediate regime. }

We study in this section the properties of the potential when the system $\bs$
 is in the high temperature phase, and the order parameter matrix
is replica symmetric and given by $Q_{a,b}=\del_{a,b}$.
According to the picture proposed in \cite{kirtir,kpv,CS} for $T_D\leq T $
this replica symmetric phase does not describe an ergodic phase,
but a situation in which
there are an exponentially large number of states, with zero overlap among
each other \cite{kirtir}.
 The triviality of the Parisi probability function $P(q)=\delta(q)$, implied
by the replica symmetric solution, is just due to entropic reasons.
It is then natural to suppose for $P$ and $R$ a
replica symmetric structure with
\beqna
P_a&=&\del_{a,1}\tp\nn\\
R_{a,b}&=&\del_{a,b}+(1-\del_{a,b})r.
\eeqna
The insertion in (\ref{25}) gives for the potential
\be
V=-{1\ov 2\B}
\l\{
2\b\b'f(\pt)-\b^2 f(r)
+\log(1-r) +
{r-\tp^2\ov 1-r}
\r\}
\ee
where $r$ is the only parameter with respect to which
the potential has to be maximized.
The maximization equations (\ref{V1},\ref{V2}) reduce to the single equation
\be
\b^2 f'(r)={r-\pt^2\ov (1-r)^2 }.
\label{eq}
\ee
It turns out that  for any concrete form of the function $f$
that we have analyzed that this equation admits a unique solution in the
range of $\b'$ where the replica symmetry is not broken.
As we stressed in the previous sections the points of extremum of $V$ with
respect to $\tp$ are of
particular importance in our analysis,
the minima correspond to stable or metastable states in which the system
can be trapped.
 Let us then write the equation that specifies
the points of extremum:
\be
\b\b' f'(\tp)={\tp\ov 1-r}
\label{ep}
\ee
We stress again that the point $r=\pt=0$ is always a solution of
(\ref{ep},\ref{eq}), and
is always an absolute minimum with $V=0$.

Let us start our  analysis considering the case of equal temperatures
$T'=T$.
In this case the qualitative features  of the potential are largely model
independent.
In figure
 1 we show the picture of the potential for four different
temperatures in the case $f(q)=q^3/2$, similar plots are obtained
for arbitrary functions.
{}From top to bottom, they represent the potential at
temperature higher then $T_D$, equal to $T_D$ between $T_D$ and $T_S$, and
right at $T_S$. We can see from the figure that for $T>T_D$
the potential is monotonically increasing, and
the only extremum of the potential is the minimum at $\tp=0$.
At the temperature $T_D$ where the dynamical transition
happens,  the potential develops for the first time  a minimum with
$\tp=r$. It is interesting to observe that
the energy in this flex point \footnote{See the following for
the definition.}  is equal to
the asymptotic value of the
 energy  in the off equilibrium dynamics.
The same is true for the parameter $r$ which turns out
to be  equal
to the dynamical Edward-Anderson parameter.

The condition for the potential of having a flex coincides with the marginality
condition.
Indeed the flex implies a zero eigenvalue in the longitudinal sector and at
$x=1$ the replicon
and the longitudinal eigenvalues are degenerate (see for example the formulae
in ref. \cite{MPR3}).
This marginality condition is well known to give exact results for the
transition temperatures
in p-spin spherical models. It also give  accurate results, compared with the
Monte Carlo, in
the case of the Ising model with random orthogonal matrix \cite{MPR3} (ROM
model).

\begin{figure}
\vspace*{23cm}
\hbox to
\hsize{ \hspace*{-4cm} \includegraphics{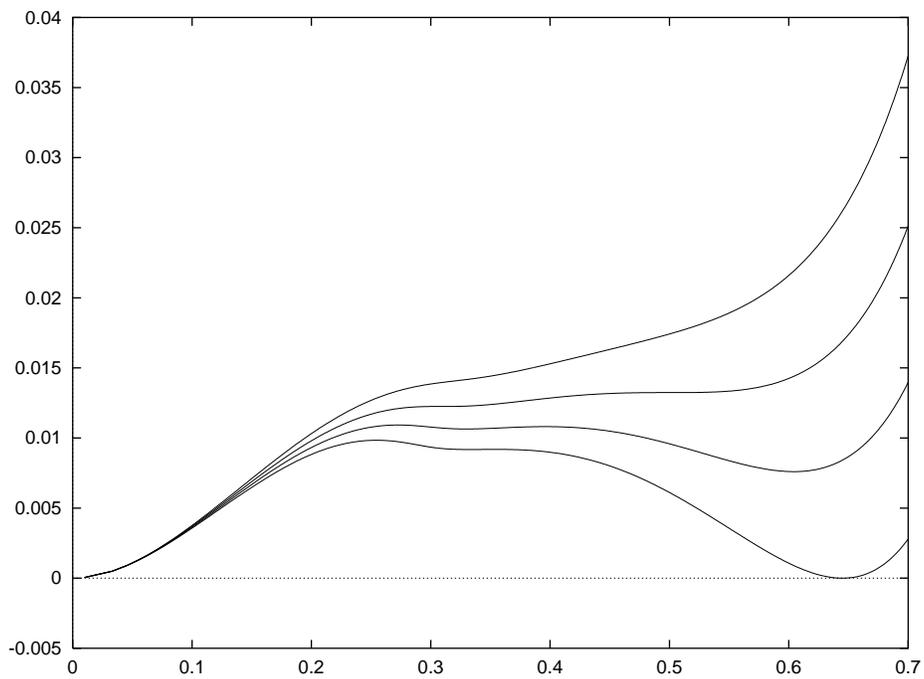} \hspace*{4cm} }
\vspace*{-15cm}
\caption{{
The potential function in the $p=3$-spin model for $T=T'$
and various temperatures. {}From top to bottom
$\b=1.59,\b=\b_D=1.63,\b=1.66,\b=\b_S=1.70,$.
Observe that a minimum is first present at the dynamical temperature, and
the value of the minimum of potential at the statical temperature is zero.
}}
\end{figure}

We have observed that in general more then one minimum can be present
in the potential. In the $p$-spin model it happens that two minima
develop at the same point.
The
rightmost one, that we will call primary is the one with  $\tp=r$, while the
other,
secondary, has $\tp<r$.
  For temperatures smaller than $T_D$ the minima have a finite depth,
i.e. are separated by extensive barriers from the absolute minimum.

The primary minimum is easily interpreted. There the system denoted
by  $s$ is in the same pure state as the system $\s$. In the region
$T_S<T<T_D$
the number of pure states is exponentially large in $N$:
${\cal N}=\e^{N \Sigma(T)}$. Consequently the probability of finding two
system in the same state is exponentially small and proportional to
$\e^{-N \Sigma(T)}$. The free energy cost to constrain two systems
to be in the same state is then proportional to the logarithm of this
probability,
namely we have
\be
V_{primary}=T\Sigma(T).
\ee
Coherently at the statical transition temperature $T=T_S$ one finds
$V_{primary}=0$.
The quantity $\Sigma$ has been computed for the $p$-spin model
in ref. \cite {CS} as the number of solution of the TAP equation
with given free energy and coincides with our calculation.
The secondary minima, could aso be associated to metastable states,
but at present we do not have  an interpretation for them.
This conclusion on the equivalence of the potential with the number of solution
of the
TAP equation hold also in the ROM \cite{PP}, and it has been argued by in ref.
\cite{remi}
on general grounds.

The study of the potential for temperatures smaller than $T_S$ would require to
take into account RSB effects, which would complicate a bit the analysis.
However it is physically clear that the shape of the potential
in that region  it is not different qualitatively from the one at $T=T_S$.
It has a minimum where $r=\pt$ are equal to the Edwards Anderson parameter
 and the value of potential is zero.

In the case of different temperatures $T\ne T'$ the properties
of the potential depend on the presence or absence of chaos with respect to
temperature changes. The primary minimum of the potential if it exists,
reflects
the properties of
the states which are of equilibrium  at temperature $T'$ when they
are followed at temperature $T$. In the $p$-spin model, where the order
of the levels do not depend on the temperature, and each level  has
temperature independent  complexity, the value of the potential
in the primary minimum can be related to the properties of the
solutions to the TAP equations when they are followed in temperature.
Denoting by $F_{TAP}(T,E')$ the free energy of the TAP states that dominate
at temperature $T'$ when they are followed at temperature $T$, we find:
\be
V_{primary}=-T\Sigma(T') + F_{TAP}(T,E')-F(T)
\ee
where $F(T)$ is the free energy at temperature $T$ and $F_{TAP}(T,E')$
is the TAP free energy at temperature $T$ of the states which are
of equilibrium at temperature $T'$. We also note that $q$ is equal
to the EA parameter of the aforementioned TAP solution.
Of particular interest is the case  $T'=T_D$ where the primary
minimum is marginally stable for any value of $T$, and has energy,
defined as $E_{primary}=(\partial / \partial \b) \b [V_{primary}-F(T)]$ and
EA parameter equal to the these of the off-equilibrium asymptotic state.
In figure 2 we show the potential for $p=3$ at fixed $\b=(\b_D+\b_S)/2$, and
various values $\b'$.
For a given $T'$ the minimum exists in the range of temperatures for which the
corresponding solutions
to the TAP equations
exist.
\begin{figure}
\vspace*{23cm}
\hbox to
\hsize{ \hspace*{-4cm} \includegraphics{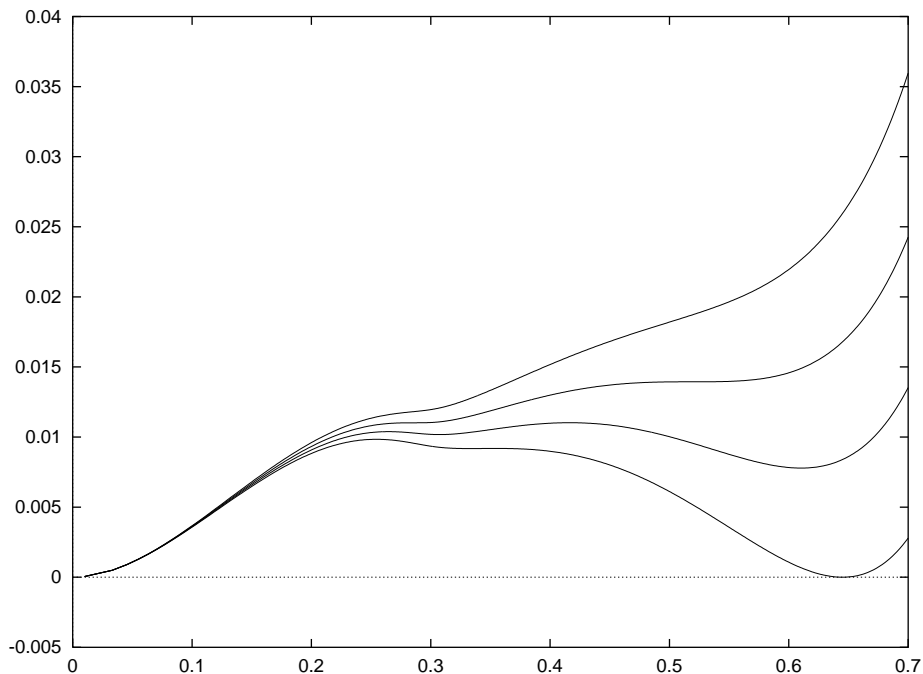} \hspace*{4cm} }
\vspace*{-15cm}
  \caption{
The potential for the $p=3$-spin model and different temperature.
The curves are drown for $\b=\b_s$ and, from top to bottom,
$\b'=1.59,\b'=\b_D=1.63,\b'=1.66,\b'=\b_S=1.70$.
}
\end{figure}

In the case models with inhomogeneous Hamiltonian
the situation  for $T\ne T'$ is different.
We did not try a systematic study of this case and we concentrated
on the case of a $3+4$ spin model with $f(q)=(1/2)(q^3+\eps q^4)$.
The first thing that we note  in this case is that the horizontal flex
 present for $T=T'=T_D$ disappears as soon as, for
$T'=T_D$ we take $T\ne T_D$.
For $T' < T_D$ it exists an interval  of values of $T$ around $T'$
in which the  minimum exists. In figure
 3 we display the potential
of the 3+4 model for fixed $T'<T_D$ and various values of $T$.
\begin{figure}
\vspace*{23cm}
\hbox to
\hsize{ \hspace*{-4cm} \includegraphics{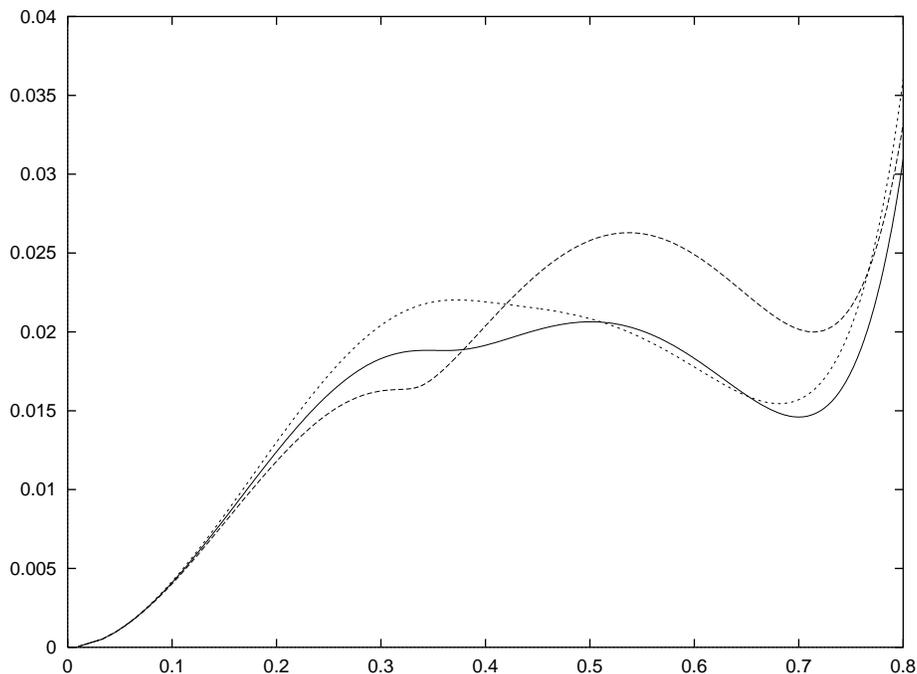} \hspace*{4cm} }
\vspace*{-15cm}
 \caption{ The potential for $f(q)=1/2 (q^3+q^4)$ for $\b'=(\b_D+\b_s)/2=1.27$
and $\b=1.21$ (dots) $\b=1.31$ (full line) and $\b=1.41$ (dashes).}
\end{figure}

One might wonder if the point of horizontal
flex of the potential are associated to the
dynamical state as it happens in the $p$-spin model. To do that we
have looked at values of $T$ and $T'$ such that
the marginality condition
$\b^2 (1-r)^2 f''(r)=0$ was verified in the minimum of the potential.
In that points we found that the minimum has a finite depth and the energy
does not coincide with that of the dynamical states.

It is tempting to interpret the minima of the potential as the
properties of  equilibrium TAP solutions at temperature $T'$ when followed
at temperature $T$, we are however very cautious on this point
due to the possibility of crossing of the solutions in free-energy.
Let us just mention that this should be valid below the
static transition   ($T'< T_S$)
where the lowest TAP states dominates the partition sum,
and where there can not be crossing.
 Chaos with respect to temperature implies that
the free energy of these solutions at temperature $T$ will be typically
higher then that of the equilibrium solution.
One could expect that  the minimum of the
potential has a value higher then zero for any $T\ne T'$. We aspect then that
$V_{primary}(T,T')\approx (T-T')^{2k}$, with integer $k$.

\section{ Dynamics}
We study now the relaxation dynamics at temperature $T$
of a system starting its evolution at time zero from an equilibrium
configuration at temperature $T'$. We start from the usual Langevin equation
for the model
\beqna
{\d s_i(t)\ov \d t}&=&-\mu(t) s_i(t)-{\partial H\ov \partial s_i}+\eta_i(t)
\nn\\
\la\eta_i(t)\eta_j(t')\ra&=&2T\del_{ij}\del(t-t')
\eeqna
As usual the time dependence of $\mu(t)$ is chosen to enforce the
spherical constraint $|\bs|^2=N$ at any time.
The dynamical generator of the correlation function \cite{MSR},
 averaged over the
initial condition is
\beqna
Z_{Dyn}=1=& &\ol{
\int \cD \eta \cD s \cD \hs
{1\ov Z(T')}\exp\{-\B'H[s(0)]\}
\cdot}
\nn\\
& &\ol{
\exp\l[
\int_0^t \d t' \sum_i i\hs_i(t')[\dot{s}_i(t')+{\partial H\ov \partial
s_i}+\mu(t)s_i(t')
+Ti\dot{s}_i(t')]
\r]
}
\label{zdyn}
\eeqna
The term $1/Z(T')$ obliges us to introduce replicas to
perform the quenched average. One possible way is to start from
the average of the logarithm of $Z_{Dyn}$ and replicate the system at each time
\cite{HJY}. We follow here a different route, using the relation
$1/Z=\lim_{n\to 0} Z^{n-1}$ and replicating only the system at time zero.
So,  ``replica number 1'' will be present at  all times,  while
replicas 2 to $n$ will only be present at time zero. We will denote
as $s_i^a(0)$ the spins of the a-th replica at time zero. Coherence
would require the notation $s_i^1(t)$ for the value of the spins at time $t$;
we will use instead $s_i(t)$.
Using standard manipulations to average over the disorder we find
\beqna
Z_{Dyn}=& &\int \cD s \cD \hs
\exp\l[
\sum_{a,b}^{1,n}\B'^2f(\bs_a(0)\cdot\bs_b(0)/N)-{\mu\ov 2}\sum_{a,i}
s_i^a(0)^2
\r]
\nn\\
& &
\exp\l[
\int_0^t \d t' \sum_i \hs_i(t')[i\dot{s}_i(t')+Ti\dot{s}_i(t')]
\r]
\nn\\
& &
\exp\l[
\int_0^t \d t'\d t''
f'(\bs(t')\cdot\bs(t'')/N)\sum_i i\hs_i(t') i\hs_i(t'') +
\r.\nn\\
& &\l.
f''(\bs(t')\cdot\bs(t'')/N)\bs(t')\cdot{i{\bf\hs}}(t'')/N
\sum_i i\hs_i(t') s_i(t'')
\r]
\nn\\
& &
\exp\l[
\b'\sum_a\int_0^t\d t' f'(\bs(t')\cdot\bs_a(0)/N)\sum_i
i\hs_i(t') s_i^a(0)
\r]
\label{pip}
\eeqna
Introducing  the correlation function
$C(t',t'')=\bs(t')\cdot\bs(t'')/N$, the response
$G(t',t'')=\bs(t')\cdot\vec{i\hs}(t'')/N$ ($t'>t''$)
and the correlation with the initial condition
$C_a(t')=\bs(t')\cdot\bs_a(0)/N$,
and observing that they are non-fluctuating quantities for $N\to\infty$,
we recognize in (\ref{pip}) the dynamical generator functional
of a system of $N$ independent spins
subject to the Langevin equation
\beqna
& &{\d s_i(t)\ov \d t}=-\mu(t) s_i(t)+\int_0^t f''(C(t,t'))G(t,t')s_i(t')+
\B'\sum_{a=1}^n f'(C_a(t))s_i^a(0)+B_i(t)
\nn\\
& &\la B_i(t)B_j(t')\ra=2T\del_{ij}\del(t-t')+\del_{i,j} f'(C(t,t'))
\eeqna
satisfying the self consistency equations:
\beqna
& &C(t,t')=\bs(t)\cdot\bs(t')/N
\nn\\
& &G(t,t')=\bs(t)\cdot i{\bf \hs}(t')/N \;\;\; t>t'
\nn\\
& &C_a(t)=\bs(t)\cdot\bs_a(0)/N.
\eeqna
It is easy to check that the ``equilibrium part'' in the  action,
involving the parameters $Q_{ab}=\bs_a(0)\cdot\bs_b(0)/N$ is
not affected by the dynamics and, as it should, describes correctly
 the equilibrium  at temperature $T'$.
{}From the Langevin equation one derives the closed system:
\beqna
{\partial G(t,t') \over \partial t}&=& -\mu(t)G(t,t')
+\int_0^t ds  \ f''(C(t,s))G(t,s)(G(t,t')-G(s,t'))\ ,
\nn\\
{\partial C(t,t') \over \partial t} &=&-\mu(t) C(t,t')
+  \int_0^{t'} ds \  f'(C(t,s)) \ G(t',s)\nonumber\\
& & + \int_0^t ds \  f''(C(t,s))G(t,s) \ (C(t,t')-C(s,t'))\
+ \b'\sum_a f'(C_a(t))C_a(t'),
\nonumber\\
{\partial C_a(t) \over \partial t}&=&-\mu(t)C_a(t)
+\int_0^t ds  \ f''(C(t,s))G(t,s) C_a(s) +\B'\sum_b f'(C_b(t))Q_{a,b}
\label{em}
 \nn\\
\mu(t)&=&
 \int_0^t ds  \ f'(C(t,s)) \ G(t,s)
+\int_0^t ds \  f''(C(t,s))G(t,s) \ (C(t,t)-C(s,t))\
\nonumber\\ & &+T \
+ \b'\sum_a f'(C_a(t))C_a(t),
 \eeqna
with the conditions:
\be
C_1(t)=C(t,0)
\;\;\;
C_a(0)=Q_{1,a}
\ee
These equations differ from the usual off-equilibrium equations for
the presence of terms which couple with the configuration at time zero.
The matrix $Q_{a,b}$ has the usual hierarchical form, and $C_a(0)$ is its first
row.
It is then natural to
assume that $C_a(t)$ continues at any time to have the structure of the
first row
of a hierarchical matrix. Once this ansatz is plugged in (\ref{em})
we get a system of equations which admits for any time a unique solution.

The basic question that can be answered  with the aid of these equations
is whether starting from equilibrium at temperature $T'$  the system
equilibrates
in a metastable state, or  it ends up in an aging state.
In this paper we will only consider non-aging solutions to the equations
(\ref{em}). We leave to future work the investigation of possible aging
solutions.

We look then for solution where no aging appears, the correlation functions
$C(t,t')$ tends for infinite times to the homogeneous function
$C_{as}(t-t')$, and the response function is related to $C_{as} $ via the
fluctuation dissipation relation:
$G_{as}(\tau)=-\B {\partial C_{as}(\tau)\ov\partial\tau}$.
 In the high temperature regime $C_a(0)=0$ and
the equations coincide with the off equilibrium ones.
  Interesting phenomena
appear for temperature $T'\leq T_D$. Let us  study the case
$T_S\leq T'\leq T_D$. Here we have $Q_{a,b}=\del_{a,b}$, a bit of reflection
on the equation for $C_a(t)$ reveals. that the replica structure of $C_a(t)$ is
at any
time of the kind
\be
C_a(t)=\del_{1,a}C(t,0).
\ee
Using then the notations:
\beqna
\tp=\lim_{t\to\infty} C(t,0)
\nn\\
r=\lim_{\tau\to\infty} C_{as}(\tau)
\nn\\
\mu_\infty=\lim_{t\to\infty} \mu (t)
\eeqna
one finds for $C_{as}(\tau)$ the equation
\be
{\partial C_{as}(\tau)\ov \partial \tau}=-\mu_\infty C_{as}(\tau)
+\B [f'(1)C_{as}(\tau)-f'(r)r]
-\B\int_0^\tau f'(C_{as}(\tau+\tau')){\partial C_{as}(\tau')\ov \partial \tau'}
+\B'\tp f'(\tp).
\ee
While $\tp,r,\mu_\infty$ are solution of the following system
\beqna
& &-\mu_\infty\tp+\B \tp\mu_\infty\B'f'(\tp)=0
\nn\\
& &-\mu_\infty r+\B q[f'(1)-f'(r)]+\B(1-r)f'(r)+\B'\tp f'(\tp)=0
\nn\\
& &-\mu_\infty +T+\B[f'(1)-f'(r)r]+\B'\tp f'(\tp)=0.
\label{K}
\eeqna
Eliminating $\mu_\infty $ we find the remarkable result that the equations
specifying $r$ coincide
with the equations (\ref{eq},\ref{ep}) of section 4
for the points of extremum of the potential.

We have seen in section 4 that depending on the form of $f$ and on the
temperatures $T,T'$
solutions to (\ref{K}) can exist or not. In the case in which such solutions
do not exist, we easily conclude that for long times an asymptotic aging regime
sets in the system.
To understand if this regime is correlated with the initial condition would
require a more
complete ansatz then the one we are using here and we leave it f
or future work. In the case solutions with $\tp\ne 0$ exist,
we  need to show that the solution to (\ref{em}) converge to one of
the solution of (\ref{K}) and to know to
which one. The problem is easily solved for equal temperatures $T=T'$. If we
start at equilibrium
we stay there. We explicitly verified that the solution to the equations
(\ref{em}) is time
translation invariant from the start. We belive  that it a general fact that
the dynamics  tends toward the primary minimum of the potential.
To verify that,  we performed the numerical integration of the system
(\ref{em}) for $f(q)=1/2 (q^3+q^4)$ with the method employed in \cite{frame}.
Figure 4
shows the approach of $C(t,0)$ to $\tp=.646$ in the case
$T=0.673$, $T'=0.804$. In the same way it is possible to see that the
energy tends to the value predicted by the potential theory.
\begin{figure}
\vspace*{23cm}
\hbox to
\hsize{ \hspace*{-4cm} \includegraphics{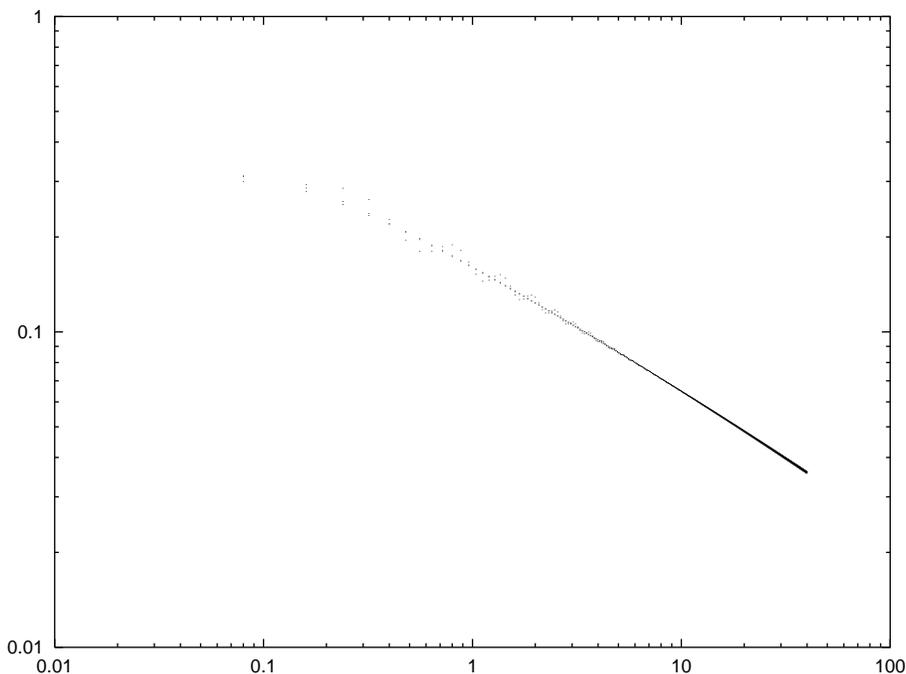} \hspace*{4cm} }
\vspace*{-15cm}
 \caption{The convergence of $C(t,0) $ to $\tp$ for the 3+4 model
in the case $T=0.673$, $T'=0.804$.   We plot
$C(t,0)-\tp$ as a function of $t$ in a log-log scale, for
$\tp$ we take the value in the primary minimum of the potential
$\tp=.646$.   $C(t,0)-\tp$ seems to approach zero as a power law.
A similar plot is found for the energy as a function of time.
}
\end{figure}

\section{Conclusions}

In this paper we have investigated the free-energy landscape
of a spherical spin-glass model, in the vicinity of equilibrium states.

The model is chosen in a way to have a low-temperature glassy phase
characterized by a single step Parisi function $q(x)$.
The study is performed introducing the `potential function'
defined as the minimal work needed to keep a system
to a given distance from an equilibrium configuration.
In this way we could identify the dynamical transition temperature as the
point where first appear a minimum in the potential function at equal
temperatures.

In the $p$-spin model, where the Hamiltonian is an homogeneous
function and there is not chaos with respect to temperature
changes, the dynamical states for $T<T_D$ can be associated to the points
where the potential has a flex. This property is not true any longer
whenever chaos is present in the model. In that case, we do not have a recipe
to associate properties of the dynamic asymptotic state to some particular
point in the phase space of our coupled system.

The minima of the potential have been associated to metastable states.
The life of the metastable states can be estimated arguing that the
free-energy barrier is just given by the maximum to minimum difference
in the potential \cite{parpot}, which is then proportional to the size of
the system $N$.
It is interesting to remind here the  result of the analysis
in \cite{parpot} in finite dimension. In that case the order parameter
is space dependent. The free energy barrier can be estimated with
instantonic techniques (see e.g. ref. \cite{parbook}), introduced in the
the contest of spin glass models in \cite{fpvinst}. It turns out that in
 three dimensions the relaxation time of  the metastable states scales
as $\tau_{max}\sim \e^{const\ov (T-T_s)^2}$, to be compared with the
Vogel-Fulcher law  $\tau_{max}\sim \e^{const\ov (T-T_s)}$, often used
to fit the temperature dependence of the viscosity near the glassy transition
in structural glasses.

Let us conclude the paper noting that simple generalizations
of the potential method would allow to gain further
insight on the free-energy landscape of spin-glass
models. For example one can consider the free energy cost
of a situation where there is a
 first replica  at equilibrium at a given temperature, a  second one,
 at another
temperature, is kept at a fixed distance from the first one, and a third
replica, at a third temperature  kept at fixed distances from the first two.

Also, it would be very interesting to generalize our calculation to
models with continuous RSB, to study the property of chaos with respect to
temperature changes in that case.

\vspace{1 cm}
{\bf Acknowledgements}

We are glad to thank L. Cugliandolo, J.Kurchan,
E. Marinari, M.M\'ezard, R.Monasson, M. Potters, F. Ritort and M.A.Virasoro
for important conversations on the topics discussed in this paper.

S.F. is grateful for warm hospitality
to the Laboratoire de Physique Th\'eorique
of the Ecole Normale Superieure in Paris, and to the
Dipartimento di Fisica dell'Universit\`a di Roma I, where
 part of this work has been elaborated.

\end{document}